\documentclass{aa}
 
\usepackage{graphicx}
\usepackage{txfonts} 

\def\Fe4{Fe~{\sc iv}}
\def\Fe17{Fe~{\sc iv}}
\newcommand{\eion}{({\rm e~+}~ion)\,}
\def\etal{{\it et\thinspace al.}\ }

\begin{document}
   \title{Atomic data from the Iron Project}
   \subtitle{LIX. New radiative transition probabilities for Fe IV 
including fine structure}
\author{S. N. Nahar and A. K. Pradhan}
   \offprints{S.\,N. Nahar}
   \institute{Department of Astronomy, The Ohio State University,
              Columbus, OH 43210, USA\\
   \email{nahar@astronomy.ohio-state.edu}
\thanks{Complete electronic data tables of energies and trans\-ition 
probabilities are available from the CDS via anonymous ftp to 
{\tt cdsarc.u-strasbg.fr} (130.79.128.5) or
via {\tt http://cdsweb.u-strasbg.fr/Abstract.html}
} }
\abstract{We present new calculations for transition probabilities 
of Fe~IV, with much more extensive datasets than heretofore available. 
The large-scale close coupling R-matrix calculations yield 1,798 LS 
bound states with n $\leq$ 11 and $l \leq$ 9, and corresponding 
transition probabilities for 138,121 dipole allowed transitions in 
the form of oscillator strengths $f$, line strengths $S$, and 
A-coeffficients for a variety of applications. This represents the 
largest R-matrix dataset in LS coupling for any ion under either the 
Opacity Project or the Iron Project. Through algebraic transformation 
of the LS multiplets, a total of 712,120 dipole allowed fine structure 
transitions for Fe IV are obtained. Observed transition energies are 
used together with the energy independent line strengths to derive 
the $f$- and the $A$-values; the adopted algorithm used calculated 
energies for the remainder. Present results show significantly better 
accuracy for the important low-lying states than previous calculations.
Monochromatic and mean opacities for Fe IV are computed and compared
with those obtained using the Opacity Project data. We find differences 
which could have important consequences for several astrophysical
applications involving low ionization stages of iron. 
}
\authorrunning{S.\,N. Nahar and A.K. Pradhan}
\titlerunning{Radiative Data for Fe IV}
\maketitle

\section{Introduction}

It is well known that low ionization stage of iron are of great 
interest for astrophysical spectroscopy and models of stellar
atmospheres (Hubeny \etal 2003), novae (McKenna \etal 1997), photoionized 
H~II regions (Dopita and Sutherland 2003, Rubin \etal 1997), and
active galactic nuclei (Sigut and Pradhan 2003, Sigut \etal 2004).
 The ion Fe~IV is of particular interest 
since its ionization potential is $\sim$ 54.8 eV,
nearly equal that of He~II at 54.4 eV. Thus, for example, in a
photoionized H~II region such as the great diffuse nebula in Orion, stellar
ionizing flux is effectively shielded by Helium to prevent extensive
ionization of Fe ions beyond Fe~IV. Whereas hot central stars of 
high-excitation planetary nebulae may well ionize beyond Fe~IV (e.g.
Hyung and Aller 1998), ordinary main sequence O, B stars have a 54 eV 
'cut-off' that attenuates the ionizing flux. Another example is the 
anomalous Fe~II emission from active galactic nuclei.
Non-local-Thermodynamic-Equilibrium (NLTE) spectral models need to
consider the first four ionization stages, Fe~I~-~IV, with Fe~IV as the
dominate stage in the fully ionized zone (e.g. Sigut \etal 2004
and references therein). In previous studies at Ohio State under the 
Iron Project, we have reported a large amount of radiative and 
collisional data for the ionization stages Fe~I-VI, including fine 
structure transition probabilities for Fe~II and Fe~III
(references at: {\it www.astronomy.ohio-state.edu/$\sim$pradhan}). The 
aim of this paper is to report similarly new calculations with improved 
accuracy for Fe~IV including fine structure that is
often required for spectral diagnostics and modeling.

However, study of Fe~IV is difficult and complexities arise due to 
its electronic structure with 5 electrons in the half-filled open 3d 
shell (ground configuration $1s^22s^22p^63s^23p^63d^5$); electron 
correlation effects for the many excited states are difficult to 
represent. Prior to the Opacity Project (OP, 1995, 1996), 
investigations on radiative processes for Fe~IV were carried 
out through atomic structure calculations. These include radiative
decay rates for forbidden transitions by  Garstang (1958), which are
available from the evaluated compilation by the National Institute 
for Standards and Technology (NIST: www.nist.gov) and Rubin and 
Froese-Fisher (1998). Fawcett (1989) has computed the data for many 
allowed transitions using a semi-empirical method due to Cowan.

Ab initio calculatons in the close coupling approximation using 
R-matrix method for the radiative processes were carried out under 
the Opacity Project (OP, 1995, 1996) by Sawey and Berrington (1992), 
and later under the Iron Project (IP, Hummer et al.  1993) by Bautista 
and Pradhan (1997). Further R-matrix calculations were reported for 
the total and state-specific electron-ion recombination rate 
coefficients using the unified treatment by Nahar, Bautista, and 
Pradhan (1998). Sawey and Berrington (1992) produced radiative data for 
photoionization and oscillator strengths for three symmetries:
sextets, quartets and doublets; these data are accessible through 
the OP database, TOPbase (Cunto \etal 1993). Later work of Bautista 
and Pradhan (1997) took more accurate account of electron-electron 
correlation effects, showing important resonant features in cross 
sections missing in the former investigation by Sawey and Berrington
(1992). However, owing to computational constraints, Bautista and 
Pradhan (1997) entirely omitted the doublet symmetry which has the maximum 
number of bound states. All previous R-matrix results are in LS coupling 
and consider only the dipole allowed $LS$ multiplets; no fine structure 
splitting has been taken into account.

Present work reports a more complete set of oscillator strengths and 
radiative decay rates of Fe IV for both the LS multiplets and allowed 
fine structure transition. Similar results have been presented earlier 
for Fe II (Nahar 1994), Fe III (Nahar and Pradhan 1996), and for some 
other ions. Present calculations consider correlation effects accurately,
as in Bautista and Pradhan (1997), and bound states of all symmetries, 
sextets, quartets, and doublets. At the present time it is
computationally prohibitive to carry out full Breit-Pauli R-matrix
calculations for fine structure transition probabilities including
relativistic effects in an ab intio manner. Among the low ionization
stages or iron, Fe~I~-~V, there has been only one such 
previous effort, for Fe V (Nahar and Pradhan 2000, Nahar et al. 2000).
challenging computationally. While efforts are under way to resolve 
the computationally challenging problems in relativistic calculations,
which are about an order of magnitude more intensive, there is
considerable need for fine structure data. We employ an algebraic
recoupling algorithm to obtain the fine structure transition
probabilities for Fe~IV, using observed energies where available, as in
previous IP works on Fe~II and Fe~III.

Finally, we employ the new radiative dataset to compute monochromatic
and mean opacities for Fe~IV, both as a test and comparison with the
earlier OP work, as well as to investigate possible effect on the
astrophysically important problem of iron opacities.

\section{Theory}

The R-matrix method as applied to the OP and IP work enables the
consideration of many excited bound states and transitions,
typically for all states with  n $\leq$ 10. General details of the
procedures employed have been described in earlier papers, such as for
Fe~II (Nahar 1993), for  obtaining oscillator strengths ($f$) and 
radiative decay rates ($A$) for dipole allowed fine structure 
transitions ($\Delta J$ = 0,$\pm$1 with $\Delta S$ = 0, even parity 
$\leftrightarrow$ odd parity) by algebraic tansformation of LS 
multiplets. The methodology of the close coupling (CC) approximation 
employing the R-matrix method for radiative processes is discussed in 
the OP papers by Seaton (1987) and Berringtion et al. (1987). The 
wavefunction $\mit\Psi(E)$ for the ($N$\,+\,1) electron system with 
total spin and orbital angular momenta symmetry $SL\pi$ or total 
angular momentum symmetry $J\pi$ is expanded in terms of `frozen' 
$N$-electron target ion functions $\chi_i$ and vector coupled collision 
electrons $\theta_i$,
\begin{equation}
\mit\Psi_E\eion = {\cal A}\sum_i \chi_i({\it ion})\theta_i
 + \sum_j c_j \Phi_j\eion,
\end{equation}
in some specific state $S_iL_i\pi_i$ or level $J_i\pi_i$, index $i$ 
marking channels $S_iL_i(J_i)\pi_i \ k_i^2\ell_i(SL\pi$~or~$ J\pi)$ 
with energy $k_i^2$ of the colliding electron described by $\theta$. 
{\bf A} is the antisymmetrization operator. The second sum expands 
correlation functions $\mit\Phi_j$ as products with $N$\,+\,1 bound 
orbital functions that (a) compensate for the orthogonality conditions 
between the continuum and the bound orbitals, and (b) represent 
additional short-range correlation that is often of crucial importance 
in scattering and radiative CC calculations for each $SL\pi$. $c_j$'s are 
variational coefficients.

{\it R}-matrix solutions of coupled equations to total symmetries $LS$
states for ($N$\,+\,1)-electron system, 
\begin{equation}
H_{N+1}\mit\Psi = E\mit\Psi,
\end{equation}
are discrete bound states $\mit\Psi_{\rm B}$ for $E\,<\,0$. Seaton 
(1985) describes the matching procedure for inner region and outer 
region wavefunctions at the R-matrix boundary which enables calculatons
of all required energy levels of the Hamiltonian matrix. 

For a bound-bound transition from an initial state $i$ to a final
state $j$, the line strength, $S$, is defined as
\begin{equation}
S = |<\Psi_j||{\bf D}||\Psi_i>|^2,
\end{equation}
in atomic units (a.u.), where the dipole operator ${\bf D}$ is
\begin{equation}
{\bf D} = \sum_n{\bf r}_n,
\end{equation}
in the length form, summed over the total number of electrons in
the ion, and $\Psi_i$ and $\Psi_j$ are the initial and the final
wavefunctions respectively. The oscillator strength $f_{ij}$ ($f$-value)
can be obtained from line strength $S$ in atomic units ss
\begin{equation}
S = (3g_i/E_{ij})f_{ij}.
\end{equation}
where $E_{ij}$ is the transition energy in rydbergs and $g_i$ is the
statistical weight factor of the initial state, with $g_i$ =
$(2S_i+1)(2L_i+1)$ for a LS multiplet, and $(2J_i+1)$ for a
fine-structure transition. The radiative transition probability, $A_{ji}$
(Einstein's A-coefficient or the radiative decay rate), can be obtained 
from the oscillator strength as
\begin{equation}
A_{ji}(a.u.) = {1\over 2}\alpha^3{g_i\over g_j}E_{ij}^2f_{ij}, 
~~~~~ A_{ji}(s^{-1}) = {A_{ji}(a.u.)\over \tau_0},
\end{equation}
where $\alpha$ is the fine-structure constant and
$\tau_o$ = 2.4191$\times 10^{-17}s$ is the atomic unit of time.

The fine-structure components of an LS multiplets can be obtained
through algebraic transformations of either line strength or directly from 
the oscillator strength (e.g. Nahar 1995). The line strength, which is 
independent of the transition energy $E_{ij}$, is a better choice. As 
the observed transition energies are determined more accurately than 
the calculated ones, use of the former with the line strengths can 
provide more accurate $f_{ij}$ and $A_{ji}$. Hence, present fine-structure
components are obtained from $S$ values whenever observed energies are 
available, and from $f$-values when calculated $E_{ij}$ are used.

The fine-structure line strengths, $S_{JJ}$, are obtained as
\begin{equation}
S_{JJ}=C(J_i,J_j)S_{LS}/[(2S_i+1)(2L_i+1)(2L_j+1)],
\end{equation}
for the allowed transitions $(\Delta J = 0,\pm 1)$. $S_i$ is the spin
which is the same as $S_j$. The values of the coefficients $C(J_i,J_j)$
can be found in Allen (1976). The $S_{JJ}$ values satisfy the condition
\begin{equation}
S_{LS} = \sum_J S_{JJ}.
\end{equation}
The fine-structure f-values, $f_{JJ}$, can be obtained directly from
$f_{LS}$ as (Seaton et al 1994): 
$$f_{JJ}(n_jS_iL_jJ_j,n_iS_iL_iJ_i) = f_{LS}(n_jS_iL_j,n_iS_iL_i)
(2J_j+1)(2L_i+1)$$ 
\begin{equation}
\times ~ W^2(L_jL_iJ_jJ_i;1S_i),
\end{equation}
where $W(L_jL_iJ_jJ_i;1S_i)$ is a Racah coefficient. The above components
also satisfy the sum rule
$$\sum_{J_iJ_j} (2J_i+1)f_{JJ}(n_jS_iL_jJ_j,n_iS_iL_iJ_i) = 
f_{LS}(n_jS_iL_j,n_iS_iL_i) ~ \times $$
\begin{equation}
(2S_i+1)(2L_i+1).
\end{equation}
The above form for fine structure components is used when one or both 
{\it LS} terms of the transition are unobserved or when not all the fine 
structure levels of an {\it LS} term are observed, and for transitions 
between high angular momentum states, such as higher than 
$H \leftrightarrow I$ for which Allen's coefficients are not available.

The lifetime of a level can be computed as
\begin{eqnarray}
\tau&=&\frac{1}{\sum_i A_{ki}}\,
\end{eqnarray}
where the sum is the total radiative transition probability for
level $k$, i.\,e.\
\begin{equation}
g_iA_{ki}^{\rm E1}=2.6774\times 10^9{\rm s}^{-1}\,(E_i-E_k)^3S^{\rm E1}(i,k)
\end{equation}
(the observed rate) in the electric dipole case. 

The opacity is a measure of radiation transport through matter. The
monochromtic opacity is a function of the photon frequency and depends 
mainly on the detailed atomic data: bound-bound oscillator strengths 
and bound-free photoionization cross sections. The quantity of 
practical interest in stellar models is the Rosseland mean opacity, 
$\kappa_R(T,\rho)$, defined as (Seaton \etal 1994)
\begin{equation}
\frac {1}{\kappa_R} = {\int_0^{\infty} \frac{1}
{\kappa_{\nu}} g(u)du},
~~~g(u) = \frac{15}{4\pi^4} u^4e^{-u}(1-e^{-u})^{-2} ,
\end{equation}
where g(u) is the Planck weighting function and $u = h\nu/kT$. 
Monochromatic opacities $\kappa_{\nu}$ depends primarily on oscillator 
strengths as
\begin{equation}
\kappa_{\nu}(i \rightarrow j) = {\pi e^2\over mc} N_i f(ij) \phi_{\nu}
\end{equation}
where $N_i$ = ion density in state i, $\phi_{\nu}$ is a profile factor,
and on photoionization cross sections, $\sigma_{PI}$, as
\begin{equation}
\kappa_{\nu} = N_i \sigma_{PI}(\nu).
\end{equation}

We compute the monochromatic and the mean opacities as a measure of the
differences between the present dataset for the bound-bound transitions
and the earlier OP data for Fe~IV.

\section{Computations}

The R-matrix computations for Fe IV are carried out using a 
wavefunction expansion of 16 terms of target Fe V belonging to the 
ground configuration, $3d^4$, as given in Table 1; the energies used
are the measured values available from the NIST compilation at 
www.nist.gov. The orbital wavefunctions of the target were obtained 
from atomic structure calculations using SUPERSTRUCTURE (Eissner et 
al. 1974), as in the earlier works of Bautista and Pradhan (1997) and 
Nahar et al. (1998). The present wavefunction expansion is a subset of 
the previous 31 term expansion, but includes singlet target terms 
which were not considered earlier. This enables the computation of the 
large number of doublet (N+1)-electron symmetries.
\begin{table}
\noindent{Table 1. Target terms and relative energies (in Ry) of Fe V 
in the wavefunction expansion of Fe IV. 
\\[.8ex]}
\begin{tabular}{rlll}
\hline
\noalign{\smallskip}
\multicolumn{3}{c}{\it Term} & \multicolumn{1}{c}{$E_{\rm o}$/Ry} \\
            \noalign{\smallskip}
\hline
            \noalign{\smallskip}
  1& $3d^4$&$^5D$& 0.000000  \\ 
  2& $3d^4$&$^3H$& 0.230232  \\
  3& $3d^4$&$^3P$& 0.234211  \\ 
  4& $3d^4$&$^3F$& 0.244942  \\
  5& $3d^4$&$^3G$& 0.274991  \\ 
  6& $3d^4$&$^1G$& 0.333399  \\
  7& $3d^4$&$^3D$& 0.334726  \\ 
  8& $3d^4$&$^1I$& 0.341832  \\ 
  9& $3d^4$&$^1S$& 0.361166 \\ 
 10& $3d^4$&$^1D$& 0.421837 \\ 
 11& $3d^4$&$^1F$& 0.480536 \\ 
 12& $3d^4$&$^3F$& 0.567719 \\ 
 13& $3d^4$&$^3P$& 0.568463 \\ 
 14& $3d^4$&$^1G$& 0.649554 \\
 15& $3d^4$&$^1D$& 0.855063 \\
 16& $3d^4$&$^1S$& 1.103820 \\ 
\hline      \noalign{\smallskip}
\end{tabular}
\end{table}

 In addition to the target expansion in Table 1, it is necessary to
ensure an accurate representation of short range correlation effects
that are vital for the bound (N+1)-electron states.
The second term of the wavefunction expansion, Eq.(1), 
consists of 61 such `bound channel' correlation functions
of Fe IV, as described by Bautista and Pradhan (1997). 

The R-matrix calculations span several stages of computations 
(Berrington \etal 1987, 1995). The computations were extensive in terms 
of CPU memory size and time requirement. Hence, they were carried out 
one or a few symmetries of $SL\pi$ at a time. The one-electron
orbital functions optimised through a configuration-interaction type
calculation for Fe V using SUPERSTRUCTURE are the input to {\tt STG1} 
of the R-matrix codes to compute Slater integrals. Radial integrals for 
the partial wave expansion in Eq.\,1 are specified for orbitals 
$0\leq\ell\leq 9$ as a basis of {\tt NRANG2 = 12} `continuum' functions 
--- sufficient for bound electrons with $n < 11$ at a radius of the 
$R$-matrix box. Computations are carried out for bound symmetries with 
0 $\leq L \leq$ 9 (both parities) and spin multiplicities, 2S+1 = 2,4,6. 

The bound energy eigenvalues are obtained from computer program known 
STGB. All bound states of the Hamiltonian matrix are scanned for up 
to $n$ = 11 and $l$ = 9 with a fine effective quantum number 
mesh of $\Delta \nu$ = 0.001. However, due to the computationl 
complexities stemming from the numerous bound states obtained,
 the code was unable to generate the standard R-matrix energy files. 
A comlementary computer code, PLSRAD, was written to sort out the 
energies directly from the eigenvalues of the Hamiltonian matrix. 

Furthermore, identification of the large number of {\it LS} bound 
states was a formidable task for this ion. With the exception of 
equivalent-electron states, the energies can be identified 
through quantum defect analysis of Rydberg series:
\begin{equation}
E = E_t - \frac{z^2}{\nu^2}\,,\qquad\quad \nu=n-\mu_{l\pm 1/2}(t)\label{eq:RyS}
\end{equation}
with series limits $E_t$ at the 16 Fe{\sc~v} `target' states, and 
the weighted percentage contributions of interacting closed channels. 
An LS term is designated according to a possible combination of the 
configuration of the core and the outer electron, and assigned an 
appropriate value of $\nu$ for the outer electron quantum numbers 
$n,l$ associted with the maximum channel percentage contribution. A 
Rydberg series is identified by sorting out the series of $\nu$ for 
the terms of same configuration but with increasing {\it n} of the 
valence electron. Code ELEVID (e.g.  Nahar 1995) was used for this 
analysis. Uncertainties in assingments of 
proper configurations are introduced for cases with small 
differences in $\nu$ for various terms of the same symmetry, as the 
quantum defects for these states are almost the same. There is also some
uncertainty in the identification of the adjacent higher or lower terms for 
such cases. The code STGBB is used to compute 
radiative data for the bound-bound transitions; the
code exploits methods developed by Seaton (1986) to evaluate the outer 
region ($>{\tt\,RA}$) contributions to the radiative transition matrix 
elements.

The f-, $S$-, and A-values for the fine structure
transitions are obtained through algebraic transformations of the LS
multiplet line strengths using spectroscopiclly observed energies.
For the cases where the observed energies are not found for all the fine
structure components of an LS term, the spitting of LS f-values
is carried out as described above using the code JJTOLS (Nahar 1995).

 Using the new bound-bound radiative dataset, but retaining the earlier OP
bound-free data (the contribution of bound-free is much smaller), we  also
calculate the monochromatic and Rosseland mean and Planck mean opacities
as described by Seaton \etal (1994). Results are compared with those
computed using the OP data calculated by Sawey and Berrington (1992).

\section{Results and Discussion}

 In the subections below we describe the results for bound state
energies of LS terms and fine structure levels levels of Fe~IV,
radiative transitions for all dipole transitions, and 
the effect of the new data on opacities. We note that this entails the
calculation and identification of the largest number of LS bound terms
(1,798) computed for any ion under the Opacity or the Iron Projects thus
far.

\subsection{Energy Levels}

A total of 1,798 bound states of Fe IV of sextet, quartet and doublet 
symmetries were obtained with n $\le \sim$ 11, $l~\le$ 9, 0 
$\leq L \leq$ 9 (both even and odd parities), and spin multiplicities 
2S+1 = 2,4,6. All energy levels have been identified and a
correspondence is established with the observed energies where available. 

In Table 2, the present bound state energies are compared with the 
measured values (available from the NIST database), the OP
calculations by Sawey and Berrington (1992), and by Bautista and Pradhan 
(1997). The low-lying energy levels in the present dataset are in very 
good agreement, within 5\%, with the measured values. However, the 
higher ones show larger differences, the highest one being about 
18\% for the state $3d^4(^1D1)4p(^2D^o)$. A reason for such larger 
uncertainty for higher states, especially the odd parity states, could 
be the improper balance for inclusion of configurations which can
improve the lower states considerably, but introduce some `spectral
repulsion' for the higher levels of the ion. Nonetheless, present 
energies on average are an improvement over the ones by Sawey and 
Berrington (1992) except for the high lying odd parity states where 
their energies show better agreement with experiment by a few percent. 
Present energies are also slightly better than those by Bautista and 
Pradhan (1997). The improvement in computed energies for the low-lying 
states is an important fact in the calculation of opacities, as 
discussed later.

\begin{table}
\noindent{Table 2. Comparison of the present energies $E_c$ of Fe IV with 
the observed binding energies $E_{\rm o}$, and computed energies, $E_{BP}$
(Bautista and Pradhan 1997) and $E_{\rm {SB}}$ (Sawey and Berrington
1992). An * next to a state indicates an incomplete multiplet with
unobserved fine structure components. 
\\[.8ex]}
\begin{tabular}{llllll}
\hline
\noalign{\smallskip}
\multicolumn{1}{c}{\it Config} & $SL\pi$ & \multicolumn{1}{c}{$E_{\rm o}$}
& \multicolumn{1}{c}{$E_{\rm c}$} & \multicolumn{1}{c}{$E_{\rm {BP}}$} &
\multicolumn{1}{c}{$E_{\rm {SB}}$}\\
 & & \multicolumn{1}{c}{(Ry)} & \multicolumn{1}{c}{(Ry)} & 
\multicolumn{1}{c}{(Ry)} & \multicolumn{1}{c}{(Ry)}\\
            \noalign{\smallskip}
\hline
            \noalign{\smallskip}
$3d5       $&    $^6S$   &   4.020000  &3.978 & 3.984 &  3.96069    \\
$3d5       $&    $^4G$   &   3.725831  &3.658 & 3.665 &  3.64011    \\
$3d5       $&    $^4P$   &   3.698270  &3.652 & 3.660 &  3.58549    \\
$3d5       $&    $^4D$   &   3.665794  &3.607 & 3.613 &  3.56341    \\
$3d5       $&    $^2I$   &   3.590930  &3.502 &       &  3.50204    \\
$3d5       $&   3$^2D$   &   3.566686  &3.509 &       &  3.43429    \\
$3d5       $&   2$^2F$   &   3.548645  &3.494 &       &  3.42470    \\
$3d5       $&    $^4F$   &   3.539584  &3.478 & 3.485 &  3.42632    \\
$3d5       $&    $^2H$   &   3.507616  &3.434 &       &  3.39455    \\
$3d5       $&   2$^2G$   &   3.495274  &3.421 &       &  3.38135    \\
$3d5       $&   1$^2F$   &   3.462191  &3.392 &       &  3.33754    \\
$3d5       $&    $^2S$   &   3.412002  &3.348 &       &  3.27170    \\
$3d5       $&   2$^2D$   &   3.344582  &3.282 &       &  3.19871    \\
$3d5       $&   1$^2G$   &   3.264596  &3.186 &       &  3.12545    \\
$3d5       $&    $^2P$   &   3.107633  &3.034 &       &  2.94101    \\
$3d5       $&   1$^2D$   &   3.033568  &2.965 &       &  2.85699    \\
$3d4(5D)4s $&    $^6D$   &   2.849206  &2.872 & 2.941 &  2.77361    \\
$3d4(5D)4s $&    $^4D$   &   2.758817  &2.778 & 2.795 &  2.68538    \\
$3d4(3P2)4s$&    $^4P$   &   2.607793  &2.619 & 2.648 &  2.53137    \\
$3d4(3H)4s $&    $^4H$   &   2.612277  &2.588 & 2.618 &  2.53129    \\
$3d4(3F2)4s$&    $^4F$   &   2.597263  &2.592 & 2.613 &  2.51606    \\
$3d4(3G)4s $&    $^4G$   &   2.569849  &2.549 & 2.580 &  2.48781    \\
$3d4(3P2)4s$&    $^2P$   &   2.552383  &2.561 & &        2.47727    \\
$3d4(3H)4s $&    $^2H$   &   2.556817  &2.530 & & \\
$3d4(3F2)4s$&    $^2F$   &   2.543005  &2.536 & & \\
$3d4(3G)4s $&    $^2G$   &   2.514627  &2.492 & & 2.43349\\
$3d4(3D)4s $&    $^4D$   &   2.510927  &2.494 & & 2.42780 \\
$3d4(1G2)4s$&    $^2G$   &   2.491359  &2.476 & & 2.40883\\        
$3d4(1I)4s $&    $^2I$   &   2.484110  &2.438 & & \\        
$3d4(1S2)4s$&    $^2S$   &   2.464203  &2.468 & & \\        
$3d4(3D)4s $&    $^2D$   &   2.458112  &2.440 & & 2.37606\\        
$3d4(1D2)4s$&    $^2D$   &   2.406789  &2.388 & & 2.32291\\        
$3d4(1F)4s $&    $^2F$   &   2.350906  &2.311 & & \\        
$3d4(5D)4p $&    $^6F^o$ &   2.295457 & 2.219& & 2.25564 \\    
$3d4(5D)4p $&    $^6P^o$ &   2.287886 & 2.213& & 2.24990 \\        
$3d4(3P1)4s$&    $^4P$   &   2.284212  &2.246 & & 2.19440 \\        
$3d4(3F1)4s$&    $^4F$   &   2.285076  &2.239 & & 2.19842\\        
$3d4(5D)4p $&    $^4P^o$ &    2.265726 & 2.183& & 2.22408\\         
$3d4(5D)4p $&    $^6D^o$ &    2.258256 & 2.174& & \\         
$3d4(3P1)4s$&    $^2P$   &    2.232085 & 2.194& & 2.14451\\         
$3d4(3F1)4s$&    $^2F$   &    2.232375 & 2.186& & 2.14682 \\         
$3d4(5D)4p $&    $^4F^o$ &    2.228839 & 2.142& & 2.18918 \\        
$3d4(1G1)4s$&    $^2G$   &    2.186593 & 2.125& & 2.09903 \\        
$3d4(5D)4p $&    $^4D^o$ &    2.176046 & 2.080& & 2.13458 \\        
$3d4(3H)4p $&    $^4H^o$ &    2.081972 & 1.977& & 2.03477 \\        
$3d4(3P2)4p$&    $^4D^o$ &    2.066669 & 1.945& & 2.02440 \\        
$3d4(3F2)4p$&    $^4G^o$ &    2.056962 & 1.940& & 2.00840 \\        
$3d4(3H)4p $&    $^4I^o$ &    2.050188 & 1.943& & 2.00550  \\        
$3d4(3P2)4p$&    $^4P^o$ &    2.039526 & 1.911& & 1.99660 \\        
$3d4(3H)4p $&    $^2G^o$ &    2.049047 & 1.937& &  \\        
$3d4(3F2)4p$&    $^4F^o$*&    2.030249 & 1.907& & 1.98676 \\        
$3d4(3H)4p $&    $^4G^o$ &    2.031609 & 1.915& & 1.98801 \\        
$3d4(3F2)4p$&    $^2D^o$*&    2.025500 & 1.899& &  \\        
$3d4(3P2)4p$&    $^2P^o$ &    2.015051 & 1.889& &  \\        
$3d4(3H)4p $&    $^2I^o$ &    2.018812 & 1.905& &  \\   
$3d4(3F2)4p$&    $^4D^o$ &    2.017690 & 1.887& & 1.96967 \\        
            \noalign{\smallskip}
 \hline
 \end{tabular}
\end{table}

The 104 observed LS term energies are obtained from statistical 
averaging of the 269 measured fine structure energies (NIST).
In contrast, the present 1.798 LS states correspond to 5,331 fine 
structure levels; with the exception of the observed energy levels, 
the fine structure transition energies are approximated as explained 
in the Theory section. 

The complete table of 1,798 bound state energies of Fe IV is available 
electronically. The energies of 104 observed states have been replaced
by measured values. The bound states have been assigned with a prefixed
index letter for convenience, as shown in sample Table 3 
(the letters may not necessarily match with those 
of the NIST table). Following NIST, it is a standard practice to 
designate the lowest
52 bound states with prefixes in alphabetical order. For even symmetries,
an ascending order of alphabet is used, first in the lower case and
then in the upper case. The opposite order is chosen for the odd parity
states, that is, descending order of alphabet, first in the lower case
and then the upper case. States lying above these 52 states do not have
any such notation; they are recognized by the energies only. 

\begin{table}
\noindent{Table 3. Sample of energies in the complete table of energies. 
\\[.8ex]}
\begin{tabular}{lll} 
\hline
\noalign{\smallskip}
\multicolumn{1}{c}{\it Config} & $SL\pi$ & \multicolumn{1}{c}{$E(Ry)$} \\
            \noalign{\smallskip}
\hline
            \noalign{\smallskip}
$3d5            $&$ a6S$&      -4.02000 \\ 
$3d4    (5D ) 4d$&$ b6S$&      -1.39828 \\
$3d4    (5D ) 5d$&$ c6S$&      -0.81636 \\
$3d4    (5D ) 6d$&$ d6S$&      -0.54091 \\
$3d4    (5D ) 7d$&$ e6S$&      -0.38568 \\
$3d4    (5D ) 8d$&$ f6S$&      -0.28891 \\
$3d4    (5D ) 9d$&$ g6S$&      -0.22449 \\
$3d4    (5D )10d$&$ h6S$&      -0.17945 \\
$3d4    (5D )11d$&$ i6S$&      -0.14672 \\
$3d4    (5D ) 4d$&$ a6P$&      -1.55722 \\
$3d4    (5D ) 5d$&$ b6P$&      -0.87599 \\
$3d4    (5D ) 6d$&$ c6P$&      -0.57173 \\
$3d4    (5D ) 7d$&$ d6P$&      -0.40367 \\
$3d4    (5D ) 8d$&$ e6P$&      -0.30030 \\
$3d4    (5D ) 9d$&$ f6P$&      -0.23216 \\
$3d4    (5D )10d$&$ g6P$&      -0.18486 \\
$3d4    (5D )11d$&$ h6P$&      -0.15068 \\
 ... &    ... &  ...  \\       
            \noalign{\smallskip}
 \hline
 \end{tabular}
\end{table}

\subsection{LS multiplets}

The LS coupling oscillator strengths are used in various astrophysical
applications, such as calculating plasma opacities. The LS multiplet 
data with 1,798 bound states of Fe IV comprises 138,121 dipole 
allowed oscillator strengths $f$ and coefficients of spontaneous 
decay, the $A-values$. The full data table is available 
electronically. Since the line strength $S$ does not depend on the 
transition energy, the present $S$ values are used to recompute the 
$f$ and $A$-values (Eq. 5) using observed energies whenever 
available to improve accuracy.

Table 4 presents a sample of the complete electronic output file, 
with the following explanation. The first line provides the ion 
information, [{\tt Z=26, \mbox{\rm number of core electrons = 22}}]. 
Note that the number of core electrons is one less than that in the 
final ion. This (somewhat confusing) notation is standard in the 
R-matrix calculations and refers to the `target ion' with which 
another electron forms a bound state. The first two columns, 
[\verb|SiLiPi(i) -> SfLfPf(f)|], refers to the symmetries of the 
initial and final bound states; the next two columns to the 
initial and final energies, $E_i$ and $E_f$, in Rydbergs, followed 
by $f$, $gf$, $A$-values. The total number of transitions is (138121) 
is also listed at the end of the headings. The $A$-values are in 
atomic units; they should be divided by 2.419$\times 10^{-17}$ to 
yield $A(sec^{-1})$. Designation for a given state in the first two 
columns consists of integers representing the spin multiplicity `S', 
which is in fact (2S+1); the total angular momentum $L$ is 0 for
S, 1 for P, etc.; the parity $Pi$ = 0 for an even state and = 1 for 
and odd state; and the fourth integer is the index of the ascending
energy order computed by STGB for the given symmetry.  

Present results for the Fe IV oscillator strengths are compared with 
earlier works in Table 5. Bautista and Pradhan (1997) considered 
transitions among sextets and quartets only and hence have relatively
smaller set of data compared to the OP data. Under the OP, Sawey and 
Berrington (1992) also produced an extensive set of data, a total of 
133,650 transitions among 1,721 bound states for all three symmetries, 
sextets, quartets, and doublets. Hence, OP data contains about 
4,500 transitions less than the present set. Fawcett (1989) employed the 
semi-empirical Cowan's Hartree-Fock relativistic code with optimized 
Slater parameters to calculate the transition probabilities of Fe IV. 
We note that the NIST database of evaluated compilation 
gives radiative decay rates only for a small number of
forbidden transitions, not allowed ones. Table 5 compares present results 
with these earlier works for some transitions. Agreement among all the
calculations is good for sextet and quartet transitions, within
10\%. However, for the doublet transitions the present oscillator strengths
and those of Sawey and Berrington (1992) show similar values, while
Fawcett's values are somewhat smaller.

\subsection{Fine structure transitions}

Fine structure $f$- and $A$-values are presented for 712,120 dipole 
allowed transitions in Fe~IV. As in the case of LS multiplets, 
observed energies of fine 
structure transitions are employed whenever available for imporved 
accuracy. As only a limited number of observed levels are available, 
a smaller set of 2,915 fine structure transitions corresponding to 
620 LS multiplets has observed energies. Rest of the transitions, 
709,205 in total, employ calculated energies; the transition 
energies are obtained as explained in the Theory section. The entire 
file containing all fine structure transitions is available 
electronically.

A sample set of $f$-, $S$-, and $A$-values for the dipole allowed fine
structure transitions in Fe IV is presented in Table 6. The first line 
of each subset corresponds to the LS transition followed by the 
fine structure components. The letter prefix designation of the
transitional states in the table correspond to their energy positions,
as explained above (see Table 3). The energies of the initial and final 
fine structure levels are given in unit of $cm^{-1}$, while 
the transitional energy differences are 
given in Rydbergs. The A-values are in $s^{-1}$. An asterisk 
(*) below an LS state indicates an incomplete set of observed energy
levels, and an asterisk for the transitional energy
indicates that one or both the levels are missing from the
observed energy set.  

Based on the accuracy of the calculated energies and comparison with 
other works, LS multiplets should be of better accuracy than those 
in TOPbase and by Bautista and Pradhan (1997). The overall accuracy 
for most of the present f-, S, and A-values should be within 20\%
for strong transitions, with the possible exception of weak 
transitions. Because of low charge of the ion, the relativistic 
effects are expected to be small.

\subsection{Opacities}

We calculate sample radiative opacities of Fe~IV for three reasons.
First, as a completeness and consistency check on the huge amount of 
data computed, second, to gauge the potential impact on iron opacities, 
and third, spectral models for determination of iron abundances in 
astrophysical sources. 

Fig. 1 shows the monochromatic opacities in most of the wavelength 
range where Fe~IV is a contributor in an astrophysical plasma
in local-thermodynamic-equilibrium (LTE). The present results are
compared with those using the bound-bound OP data by Sawey and
Berrington (1992). In order to compare only the main source of 
opacity (lines), we have used the same bound-free (photoionization) 
data in both calculations; the bound-free opacity is less than 10\% 
of the total. We choose the temperature log(T) = 4.5 and the electron 
density $N_e$ = 17.0. The Mihalas-Hummer-Dappen equation-of-state 
(MHD-EOS; Mihalas \etal 1998) used in the OP work gives
relatively large value of the ionization fraction Fe~IV/Fe = 0.57;
Fe~IV dominates the iron opacity at this temperature and density. 
Fig. 1 shows immediately the reason that Fe~IV is the dominant ion 
in the UV, between 500-3000 \AA. It is noteworthy that the
opacity of Fe~IV varies over nearly 10 orders of magnitude, implying
that it remains a large contributor over a very wide range of radiation.

The two datasets displayed in Fig. 1 differ considerably. The main 
features throughout the oapcity spectrum appear at different
positions. For example, there is relative displacement of nearly 
200 \AA in the position of the trough which is at $\sim$1500 \AA in 
the present results, but at $\sim1700$ \AA using the OP data. Even 
more striking are the diffrences in the present and the OP Rosseland 
mean opacities, 3.3 and 4.87 cm$^2$/g respectively. This nearly 
50\% difference is similar to the one we found earlier in the Fe~II 
opacities (Nahar and Pradhan 1994) using new IP data as opposed to 
the OP data by Sawey and Berrington (1997), which has been shown to 
be inaccurate for low ionization states of iron Fe~I-~IV (Nahar and 
Pradhan 1994, Bautista and Pradhan 1997, Cowley and Bautista 2003). 
We note that the calculation of latest stellar opacities under the OP
has been carried out using the OP data as well as data from
other sources (M.J. Seaton, private communication).

Although more extensive investigations are needed to ascertain the 
full range of differences at all temperatures and densities, we 
analyze the results somewhat further. The improvement in computed 
energies in the present calculations for the low-lying states is 
an important fact in the calculation of opacities, since they 
account for most of the level population; higher levels are (a) 
not populated significantly, and (b) smeared out or dissolved due 
to stark and electron impact broadening as incorporated in MHD-EOS. 
An examination of the all level populations of Fe~IV
at T = 3.16 $\times$ 10$^5$ K and N$_e$ = 10$^{17}$ cm$^{-3}$,
obtained from the MHD-EOS shows that there is nearly a sharp drop in
population of levels as one approaches the highest of the 16 ground
configuration 3d$^5$ levels. For example, the population drops by 
about 2 orders of magnitude from the ground state $3d^5(^6S)$ at 
$\sim$ -4 Ryd, to the first of the excited configuration
state $3d^44s (^6D)$ at $\sim$ -2.8 Ryd. Even higher levels do 
not contribute to the opacity in any significant way. The decrease 
is related to the negative exponential functional form in
the modified Boltzmann-Saha equations in MHD-EOS.

\section{Conclusions}

Results are presented from a large and extensive calculation for 
Fe IV. All 1,798 states were identified in spectroscopic notation 
through quantum defect and channel contribution analysis. However, 
level identifications for highly mixed excited states should be 
treated with caution since often no definitive assignment is 
possible, and in any case not needed for practical applications. 
Present energies agree very well with the observed enegies for all 
low lying terms. However, the uncertainty increases for excited 
states with higher energies, especially the odd parity states. A 
small subset of the total number of corresponding 138,121 
transitions among the LS terms, and 712,120 fine structure 
transitions, was reprocessed wherever observed energies are available. 
While the transition probabilities for the strong transitions 
should not exceed about 20\%, the uncertainty in the weak fine 
structure compoments could be much higher, especially if relativistic 
effects are important and deviations from LS coupling are significant.

The new Fe~IV results should be particularly useful for the analysis 
and application to ultraviolet to optical spectra from astrophysical 
and laboratory sources. All data files are electronically available 
from the CDS.

%
\begin{acknowledgements}
This work was partially supported by NASA (SNN) and U.S. National 
Science Foundation (AKP). The computational work was largely carried 
out on the Cray SV1 at the Ohio Supercomputer Center in Columbus, Ohio.
\end{acknowledgements}

\clearpage

\onecolumn
\begin{table*}
\noindent{Table 4. Oscillator strengths, $f$, $gf$-values and radiative
decay rates, $A$ for transitions in Fe IV [{\tt Z=26, \mbox{\rm number
of core electrons=22}}]. The first transition, [\verb|6 0 0 1  6 1 1 1|], 
means $^6S(1) \rightarrow ^6P^o(1)$, where the number within parenthesis
is the energy position of the state. The energies are Rydberg and the
$A$-values are in atomic units, i.e. $A(sec^{-1} = A(a.u.)/2.419 \times
10^{-17}$.}
\begin{verbatim}
 ______________________________________________________________________________
   26   22

  SiLiPi(i) -> SfLfPf(f)    Ei         Ef        fif         gfl        Afi
 138121       
  6  0 0  1    6  1 1  1  -4.020000  -2.287886  4.2259E-01  2.536E+00 3.395E+09
  6  0 0  1    6  1 1  2  -4.020000  -1.163950  4.8711E-02  2.923E-01 1.064E+09
  6  0 0  1    6  1 1  3  -4.020000  -1.017630  4.3453E-01  2.607E+00 1.049E+10
  6  0 0  1    6  1 1  4  -4.020000  -0.717120  1.7706E-02  1.062E-01 5.172E+08
  6  0 0  1    6  1 1  5  -4.020000  -0.650580  2.3136E-01  1.388E+00 7.032E+09
  6  0 0  1    6  1 1  6  -4.020000  -0.487200  8.8916E-03  5.335E-02 2.971E+08
  6  0 0  1    6  1 1  7  -4.020000  -0.451150  1.2883E-01  7.730E-01 4.393E+09
  6  0 0  1    6  1 1  8  -4.020000  -0.352670  5.1805E-03  3.108E-02 1.865E+08
  6  0 0  1    6  1 1  9  -4.020000  -0.331080  7.7609E-02  4.657E-01 2.828E+09
  6  0 0  1    6  1 1 10  -4.020000  -0.267150  3.3135E-03  1.988E-02 1.249E+08
  6  0 0  1    6  1 1 11  -4.020000  -0.253220  4.9970E-02  2.998E-01 1.898E+09
  6  0 0  1    6  1 1 12  -4.020000  -0.209400  2.2580E-03  1.355E-02 8.778E+07
  6  0 0  1    6  1 1 13  -4.020000  -0.199880  3.3953E-02  2.037E-01 1.327E+09
  6  0 0  1    6  1 1 14  -4.020000  -0.168560  1.6124E-03  9.675E-03 6.404E+07
  6  0 0  1    6  1 1 15  -4.020000  -0.161760  2.4110E-02  1.447E-01 9.609E+08
  6  0 0  1    6  1 1 16  -4.020000  -0.138610  1.1931E-03  7.159E-03 4.812E+07
  6  0 0  1    6  1 1 17  -4.020000  -0.133580  1.7723E-02  1.063E-01 7.167E+08
  6  1 1  1    6  0 0  2  -2.287886  -1.398280  2.1035E-01  3.786E+00 4.011E+09
  6  0 0  2    6  1 1  2  -1.398280  -1.163950  3.2583E-01  1.955E+00 4.792E+07
  6  0 0  2    6  1 1  3  -1.398280  -1.017630  1.0472E+00  6.283E+00 4.062E+08
  6  0 0  2    6  1 1  4  -1.398280  -0.717120  5.6950E-03  3.417E-02 7.075E+06
  6  0 0  2    6  1 1  5  -1.398280  -0.650580  3.3867E-03  2.032E-02 5.069E+06
  6  0 0  2    6  1 1  6  -1.398280  -0.487200  1.4603E-03  8.762E-03 3.246E+06
  6  0 0  2    6  1 1  7  -1.398280  -0.451150  1.1753E-02  7.052E-02 2.823E+07
  6  0 0  2    6  1 1  8  -1.398280  -0.352670  6.6550E-04  3.993E-03 1.948E+06
  6  0 0  2    6  1 1  9  -1.398280  -0.331080  9.9583E-03  5.975E-02 3.037E+07
  6  0 0  2    6  1 1 10  -1.398280  -0.267150  3.7717E-04  2.263E-03 1.292E+06
  6  0 0  2    6  1 1 11  -1.398280  -0.253220  7.4100E-03  4.446E-02 2.601E+07
  6  0 0  2    6  1 1 12  -1.398280  -0.209400  2.3900E-04  1.434E-03 9.045E+05
  6  0 0  2    6  1 1 13  -1.398280  -0.199880  5.4667E-03  3.280E-02 2.102E+07
  6  0 0  2    6  1 1 14  -1.398280  -0.168560  1.6250E-04  9.750E-04 6.579E+05
  6  0 0  2    6  1 1 15  -1.398280  -0.161760  4.0933E-03  2.456E-02 1.676E+07
  6  0 0  2    6  1 1 16  -1.398280  -0.138610  1.1605E-04  6.963E-04 4.930E+05
  6  0 0  2    6  1 1 17  -1.398280  -0.133580  3.1250E-03  1.875E-02 1.339E+07
  ............................................................................
 ______________________________________________________________________________
\end{verbatim}
\end{table*}

\begin{table*}
\noindent{Table 5. Comparison of present $f$-values for Fe IV with 
previous works: a - Bautista and Pradhan (1997), b - Sawey and 
Berrington (1992), c - Fawsett (1989).
\\[.8ex]}
\begin{tabular}{l@{\hspace{2.4em}}llllrl}
\hline
\noalign{\smallskip}
\multicolumn{1}{c}{Present} & \multicolumn{1}{c}{Others} &
\multicolumn{1}{c}{$C_i~-~C_k$} & $SL_i-SL_k$ & \multicolumn{1}{c}{$J_i-~J_k$}
 \\
            \noalign{\smallskip}
\hline
            \noalign{\smallskip}
0.1817 & 0.2037$^a$,0.1973$^b$,0.2037$^c$ & $3d^4(^5D)4s-3d^4(^5D)4p$ & 
  $^6D-^6P^o$ & 30 - 18 \\
0.3085 & 0.339$^a$,0.3277$^b$,0.2830$^c$& $3d^4(^5D)4s-3d^4(^5D)4p$ & 
  $^6D-^6D^o$ & 30 - 30 \\
0.4142 & 0.4603$^a$,0.4433$^b$,0.4633$^c$& $3d^4(^5D)4s-3d^4(^5D)4p$ & 
  $^6D-^6F^o$ & 30 - 42 \\
0.1710 & 0.1883$^a$,0.1820$^b$,0.1100$^c$& $3d^4(^5D)4s-3d^4(^5D)4p$ & 
  $^4D-^4P^o$ & 20 - 12 \\
0.3165 & 0.351$^a$,0.3335$^b$,0.3285$^c$& $3d^4(^5D)4s-3d^4(^5D)4p$ & 
  $^4D-^4D^o$ & 20 - 20 \\
0.4164 & 0.4583$^a$,0.4395$^b$,0.4375$^c$& $3d^4(^5D)4s-3d^4(^5D)4p$ & 
  $^4D-^4F^o$ & 30 - 28 \\
0.1599 & 0.1491$^b$,0.1073$^c$& $3d^4(^5H)4s-3d^4(^5H)4p$ & $^2H-^2G^o$ 
  & 22 - 18 \\
0.2679 & 0.2559$^b$,0.2018$^c$& $3d^4(^5H)4s-3d^4(^5H)4p$ & $^2H-^2H^o$ & 
  22 - 22 \\
0.3472 & 0.3645$^b$,0.3136$^c$& $3d^4(^5H)4s-3d^4(^5H)4p$ & $^2H-^2I^o$ & 
  22 - 26 \\
            \noalign{\smallskip}
\hline
 \end{tabular}
\end{table*}

\clearpage

\begin{table*}
\noindent{Table 6. Sample data for allowed fine structure transitions 
of Fe IV. }\\ 
\begin{tabular}{llllrrlll}
\hline
\noalign{\smallskip}
\multicolumn{1}{l}{$SL_i-SL_k$} & \multicolumn{1}{c}{$E_i$} &
\multicolumn{1}{c}{$E_k$} & \multicolumn{1}{c}{$E_{ik}$} & 
\multicolumn{1}{c}{$g_i$} & \multicolumn{1}{c}{$g_k$} & 
\multicolumn{1}{c}{$f_{ik}$} & \multicolumn{1}{c}{S}&\multicolumn{1}{c}{A} \\
& \multicolumn{1}{c}{$Ry/cm^{-1}$} & \multicolumn{1}{c}{$Ry/cm^{-1}$} &
\multicolumn{1}{c}{$Ry/\AA$} & & & & & \multicolumn{1}{c}{$s^{-1}$} \\
            \noalign{\smallskip}
\hline
            \noalign{\smallskip}
 $a~^6S^e\rightarrow z~^6P^o$ &  4.0200 &  2.2879 &  1.732E+00 &    6 &   18 &
 4.226E-01 &  4.392E+00 &  3.395E+09 \\
 &      0.000 &  190226.00 &  5.257E+01 &    6 &    8 &   1.880E-01 &  1.952E+00
 &  3.403E+09 \\
 &      0.000 &  190008.00 &  5.263E+01 &    6 &    6 &   1.408E-01 &  1.464E+00
 &  3.391E+09 \\
 &      0.000 &  189885.00 &  5.266E+01 &    6 &    4 &   9.381E-02 &  9.759E-01
 &  3.384E+09 \\
 $a~^6D^e\rightarrow z~^6P^o$ &  2.8492 &  2.2879 &  5.613E-01 &   30 &   18 &
 1.827E-01 &  2.929E+01 &  7.704E+08 \\
 & 128967.000 &  190226.00 &  1.632E+02 &   10 &    8 &   1.817E-01 &  9.762E+00
 &  5.684E+08 \\
 & 128541.000 &  190226.00 &  1.621E+02 &    8 &    8 &   6.554E-02 &  2.798E+00
 &  1.664E+08 \\
 & 128191.000 &  190226.00 &  1.612E+02 &    6 &    8 &   1.461E-02 &  4.653E-01
 &  2.813E+07 \\
 & 128541.000 &  190008.00 &  1.627E+02 &    8 &    6 &   1.170E-01 &  5.011E+00
 &  3.930E+08 \\
 & 128191.000 &  190008.00 &  1.618E+02 &    6 &    6 &   1.120E-01 &  3.579E+00
 &  2.855E+08 \\
 & 127929.000 &  190008.00 &  1.611E+02 &    4 &    6 &   5.523E-02 &  1.171E+00
 &  9.464E+07 \\
 & 128191.000 &  189885.00 &  1.621E+02 &    6 &    4 &   5.692E-02 &  1.822E+00
 &  2.167E+08 \\
 & 127929.000 &  189885.00 &  1.614E+02 &    4 &    4 &   1.286E-01 &  2.733E+00
 &  3.293E+08 \\
 & 127766.000 &  189885.00 &  1.610E+02 &    2 &    4 &   1.842E-01 &  1.952E+00
 &  2.371E+08 \\
 $a~^6D^e\rightarrow z~^6D^o$ &  2.8492 &  2.2583 &  5.910E-01 &   30 &   30 &
 3.085E-01 &  4.698E+01 &  8.654E+08 \\
 & 128967.000 &  193789.00 &  1.543E+02 &   10 &   10 &   2.510E-01 &  1.275E+01
 &  7.035E+08 \\
 & 128967.000 &  193386.00 &  1.552E+02 &   10 &    8 &   5.669E-02 &  2.897E+00
 &  1.962E+08 \\
 & 128541.000 &  193789.00 &  1.533E+02 &    8 &   10 &   7.178E-02 &  2.897E+00
 &  1.631E+08 \\
 & 128541.000 &  193386.00 &  1.542E+02 &    8 &    8 &   1.380E-01 &  5.607E+00
 &  3.872E+08 \\
 & 128541.000 &  192595.00 &  1.561E+02 &    8 &    6 &   9.789E-02 &  4.025E+00
 &  3.572E+08 \\
 & 128541.000 &  192595.00 &  1.561E+02 &    8 &    6 &   9.789E-02 &  4.025E+00
 &  3.572E+08 \\
 & 128191.000 &  193386.00 &  1.534E+02 &    6 &    8 &   1.328E-01 &  4.025E+00
 &  2.825E+08 \\
 & 128191.000 &  192595.00 &  1.553E+02 &    6 &    6 &   5.249E-02 &  1.610E+00
 &  1.452E+08 \\
 & 128191.000 &  193271.00 &  1.537E+02 &    6 &    4 &   1.238E-01 &  3.759E+00
 &  5.248E+08 \\
 & 127929.000 &  192595.00 &  1.546E+02 &    4 &    6 &   1.846E-01 &  3.759E+00
 &  3.432E+08 \\
 & 127929.000 &  193271.00 &  1.530E+02 &    4 &    4 &   3.419E-03 &  6.891E-02
 &  9.738E+06 \\
 & 127929.000 &  193120.00 &  1.534E+02 &    4 &    2 &   1.210E-01 &  2.443E+00 &  6.857E+08 \\
 & 127766.000 &  193271.00 &  1.527E+02 &    2 &    4 &   2.431E-01 &  2.443E+00
 &  3.478E+08 \\
 & 127766.000 &  193120.00 &  1.530E+02 &    2 &    2 &   6.933E-02 &  6.985E-01
 &  1.975E+08 \\
$a~^6D^e\rightarrow z~^6F^o$ &  2.8492 &  2.2955 &  5.537E-01 &   30 &   42 &
4.142E-01 &  6.731E+01 &  7.286E+08 \\
& 128967.000 &  190276.00 &  1.631E+02 &   10 &   12 &   3.582E-01 &  1.923E+01
&  7.483E+08 \\
& 128967.000 &  189515.00 &  1.652E+02 &   10 &   10 &   5.424E-02 &  2.949E+00
&  1.326E+08 \\
& 128967.000 &  188904.00 &  1.668E+02 &   10 &    8 &   4.319E-03 &  2.372E-01
&  1.293E+07 \\
& 128541.000 &  189515.00 &  1.640E+02 &    8 &   10 &   3.020E-01 &  1.305E+01
&  5.992E+08 \\
& 128541.000 &  188904.00 &  1.657E+02 &    8 &    8 &   9.991E-02 &  4.359E+00
&  2.428E+08 \\
& 128541.000 &  188428.00 &  1.670E+02 &    8 &    6 &   1.246E-02 &  5.481E-01
&  3.975E+07 \\
& 128191.000 &  188904.00 &  1.647E+02 &    6 &    8 &   2.542E-01 &  8.270E+00
&  4.687E+08 \\
& 128191.000 &  188428.00 &  1.660E+02 &    6 &    6 &   1.359E-01 &  4.456E+00
&  3.288E+08 \\
& 128191.000 &  188086.00 &  1.670E+02 &    6 &    4 &   2.333E-02 &  7.693E-01
&  8.373E+07 \\
& 127929.000 &  188428.00 &  1.653E+02 &    4 &    6 &   2.121E-01 &  4.616E+00
&  3.451E+08 \\
& 127929.000 &  188086.00 &  1.662E+02 &    4 &    4 &   1.669E-01 &  3.654E+00
&  4.029E+08 \\
& 127929.000 &  187878.00 &  1.668E+02 &    4 &    2 &   3.240E-02 &  7.116E-01
&  1.553E+08 \\
& 127766.000 &  188086.00 &  1.658E+02 &    2 &    4 &   1.821E-01 &  1.987E+00
&  2.209E+08 \\
& 127766.000 &  187878.00 &  1.664E+02 &    2 &    2 &   2.283E-01 &  2.500E+00
&  5.502E+08 \\
 $a~^4P^e\rightarrow z~^4S^o$ &  3.6983 &  1.6735 &  2.025E+00 &   12 &    4 &
 6.310E-02 &  1.122E+00 &  6.233E+09 \\
 &  35253.800 &  257503.00 &  4.499E+01 &    6 &    4 &   6.311E-02 &  5.609E-01
 &  3.119E+09 \\
 &  35333.300 &  257503.00 &  4.501E+01 &    4 &    4 &   6.309E-02 &  3.739E-01
 &  2.077E+09 \\
 &  35406.600 &  257503.00 &  4.503E+01 &    2 &    4 &   6.307E-02 &  1.870E-01 &  1.038E+09 \\
 $a~^4P^e\rightarrow z~^4P^o$ &  3.6983 &  2.2657 &  1.433E+00 &   12 &   12 &
 5.704E-02 &  1.433E+00 &  9.402E+08 \\
 &  35253.800 &  193549.00 &  6.317E+01 &    6 &    6 &   4.020E-02 &  5.017E-01
 &  6.720E+08 \\
 &  35253.800 &  191694.00 &  6.392E+01 &    6 &    4 &   1.703E-02 &  2.150E-01
 &  4.170E+08 \\
 &  35333.300 &  193549.00 &  6.320E+01 &    4 &    6 &   2.583E-02 &  2.150E-01
 &  2.875E+08 \\
 &  35333.300 &  191694.00 &  6.395E+01 &    4 &    4 &   7.564E-03 &  6.371E-02
 &  1.234E+08 \\
 &  35333.300 &  191021.00 &  6.423E+01 &    4 &    2 &   2.354E-02 &  1.991E-01
 &  7.611E+08 \\
 &  35406.600 &  191694.00 &  6.398E+01 &    2 &    4 &   4.726E-02 &  1.991E-01
 &  3.849E+08 \\
 &  35406.600 &  191021.00 &  6.426E+01 &    2 &    2 &   9.410E-03 &  3.982E-02
 &  1.520E+08 \\
            \noalign{\smallskip}
 \hline
 \end{tabular}
\end{table*}

   \begin{figure}
   \centering
   \includegraphics[width=\textwidth]{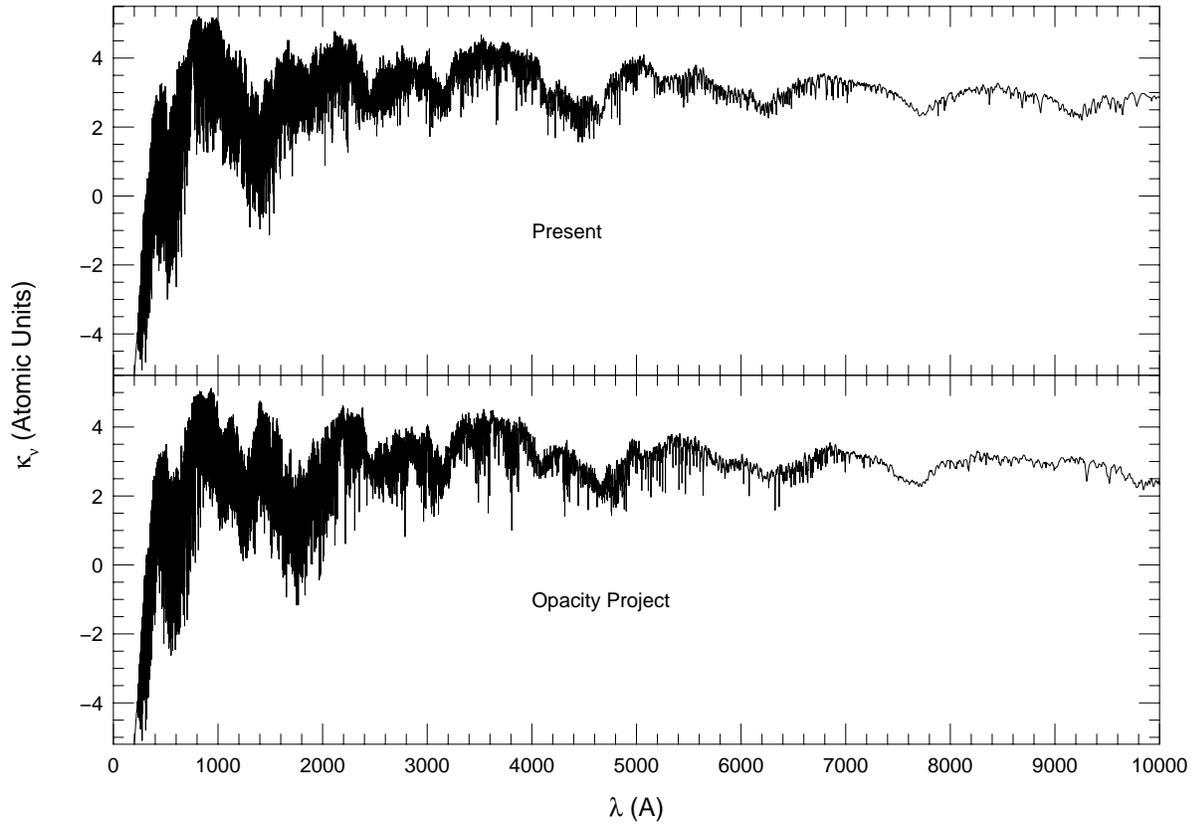}
      \caption{Monochromatic opacities of Fe IV using the present data (top
panel) and with the Opacity Project data (lower panel) from Sawey and 
Berrington (1992) at Log T  (K) = 4.5 and Log N$_e$ (cm$^{-3}$) = 17.0. 
The major features show a shift in wavelength or differences in intensities.}
         \label{FigVibStab}
   \end{figure}

\end{document}